# Climbing the N-shell resonance ladder of xenon

Steffen Palutke[1], Michael Martins[2], Stephan Klumpp[1], Karolin Baev[1,2], Mathias Richter[3], Tobias Wagner[2], Marion Kuhlmann[1], Mabel Ruiz-Lopez[1], Michael Meyer[4] and Kai Tiedtke[1]

[1]Deutsches Elektronen-Synchrotron DESY, Notkestr. 85, 22607 Hamburg, Germany
[2]Department of Physics, University of Hamburg, Luruper Chaussee 149, 22761 Hamburg, Germany
[3]Physikalisch-Technische Bundesanstalt, Abbestr. 2-12, 10587 Berlin, Germany
[4]European XFEL, Holzkoppel 4, 22869 Schenefeld, Germany

The dependency on the excitation energy of ultrafast multi-photon ionization of xenon by intense, short extreme ultraviolet pulses (XUV) was investigated in the vicinity of the $4d$ 'giant' resonance using ion time-of-flight spectroscopy. The yields of the high charge states of xenon show strong variations with the excitation energy. With reference to simulated absorption spectra, we can link the photon energy dependency to resonance structures of single-electron excitations mainly in the xenon N-shell and purely sequential multi-photon absorption.

## I. Introduction

Since the early days of spectroscopy, rare gases are widely used, either as commissioning and calibration sample of spectrometers or as part of pioneering experiments whenever new light sources become available, such as free-electron lasers (FELs).

Free-electron lasers allow the generation of intense short light pulses from the XUV to the hard x-ray regime. With up to $10^{13}$ photons per pulse with pulse durations from a few fs to a few 100 fs, FELs generate peak brilliances which are up to ten orders of magnitude higher than those of modern third-generation radiation sources. FELs enable the investigation of the interaction of matter with extreme laser fields and atomic and electronic dynamics on fs time scales. Pioneering experiments have been performed in the XUV and x-ray regime mostly on the aforementioned rare-gas atoms, but also on molecules and clusters [1-7].

Despite being seemingly simple systems, rare gases, especially xenon, generate up to now new insides into multi-photon ionization and correlated processes induced by the short XUV and x-ray pulses, see for example [3, 5, 8-13].

The electron-rich xenon atom acts as a model system to pave the way to understand the ionization and relaxation processes in more complex electronic systems such as molecules and proteins which contain heavy atoms. An important specific example is iodine, which is a trace element with a vital impact on atmospheric chemistry [14] and biological processes in living organisms [15]. One of the pioneering experiments was performed by Sorokin et al. [3] on xenon using FEL pulses in the XUV region. They found $Xe^{21+}$ ions after the irradiation with highly intense XUV pulses in the vicinity of the xenon $4d$ 'giant' resonance [16, 17] at a photon energy of about 93 eV (13.3 nm). According to this work, at least 57 photons have to be absorbed in total by a xenon atom to be ionized 21-fold. Seven XUV photons alone are required at least to overcome the $Xe^{20+}$ ionization threshold from its ground state.

Under so far considered experimental conditions, it is established that up to a charge state of $Xe^{5+}$ the ionization can be explained by sequential one-photon absorption steps followed by Auger-Meitner processes or direct photoelectron emission. Starting from $Xe^{5+}$ multi-photon processes play a role. [18] However, even at fluences typically realized at FELs, the simultaneous absorption of up to seven XUV photons is very unlikely. Therefore, the emergence of $Xe^{21+}$ at 93.2 eV photon energy was discussed intensively and controversially [19-21], but the underlying mechanisms are still not conclusively clarified.

In order to shed light on this question, efficient photon energy tunability and high levels of irradiance are required in combination with instrumentation that allows for collecting all created ions with high accuracy and mass-to-charge resolution.

For this, we performed excitation energy-dependent ion time-of-flight (iToF) spectroscopy of xenon ions interacting with short XUV FEL pulses. We found a very strong dependency of the yield of highly charged ions on the excitation energy.

To explain this, we present a possible mechanism to achieve very high charge states in xenon by multiple resonant sequential single-photon absorption processes via intermediate resonances. This mechanism mainly involves single-electron excitations within the xenon N-shell.

## II. Experiment

The experiment has been performed at the FL24 beamline of the free-electron laser facility FLASH2 at DESY in Hamburg, Germany, which fulfills the aforementioned requirements. FLASH2 is equipped with variable gap undulators, which enable changing the FEL photon energy within minutes [22]. In addition, the FL24 beamline is equipped with a set of bendable Kirkpatrick–Baez mirror optics to adjust the focus during the experiment [23]. To excite the neutral xenon, 13 different photon energies ranging from 82.6 eV up to 126.5 eV have been chosen in the region of the xenon 4$d$ 'giant' resonance [16, 17]. The photon energy bandwidth of a self-amplified spontaneous emission FEL [24], such as FLASH, is typically about 1%.

Ion time-of-flight mass spectra of the excited xenon have been recorded with a shot-to-shot scheme operating FLASH2 in single-bunch mode. The interaction volume was located between a pusher plate set to negative voltage and an iToF spectrometer working in a Wiley-McLaren configuration [25]. The pulse intensity of the delivered XUV pulses was determined by a gas monitor detector (GMD) [26] in front of the beamline as part of the beamline infrastructure and behind the spectroscopic chamber with a second GMD. With thin metal filters, the average pulse energy was varied between 3 µJ and 25 µJ. The vertical and horizontal diameter of the FEL beam focus was determined by wavefront measurements using a Hartmann-type setup [27] for each individual FEL photon energy and ranged from 3.5 µm to 7.0 µm full width at half maximum (FWHM). From the pulse energy and spot size, the fluence of the XUV pulses in the focal volume of the iToF spectrometer was calculated in J/cm². All measured spectra have been sorted according to the measured pulse fluence value and binned with a bin size of 2 J/cm² per bin in the fluence range up to 100 J/cm². The FWHM pulse duration was estimated from electron bunch parameters to be about 150±50 fs. The FEL photon energy has been measured via an online electron spectroscopy-based diagnostics setup [28].

## III. Results

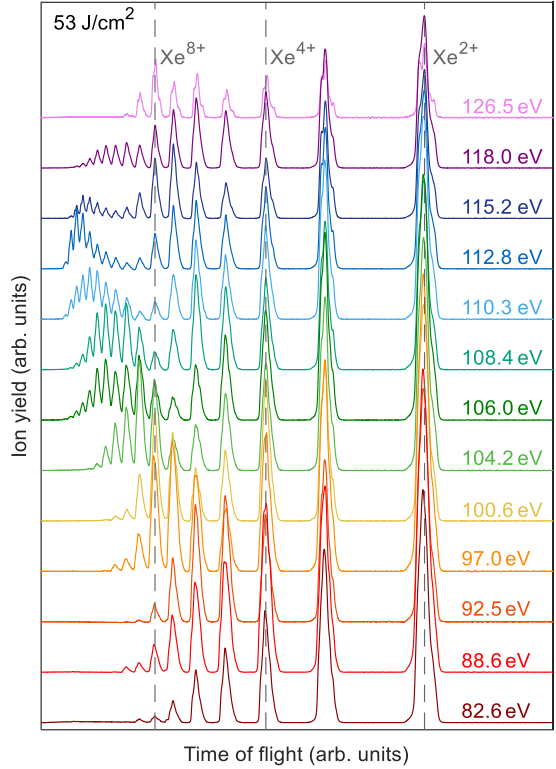

**Figure 1:** Excitation energy dependent photoionization spectra of xenon at a pulse fluence of 53 J/cm² normalized by the respective number of single-shot spectra.

Figure 1 shows the ion yield spectra of xenon at different excitation energies from 82.6 eV to 126.5 eV (15.0 nm to 9.8 nm) at 53 J/cm² (∼3.5×10$^{18}$ photons/cm²). Charged states of up to Xe$^{19+}$ are created for photon energies close to the maximum of the giant resonance at around 112 eV photon energy. Upon excitation the wings of the giant resonance, the highest charge state observed is around Xe$^{10+}$. All in all, the spectra show strong changes in the ion yield production by XUV pulses for the chosen different photon energies. Every spectrum was normalized by the number of single-shot spectra measured for the respective beam parameters.

By fitting the observed charge state peaks in Figure 1 with multi-Gaussian profiles and plotting the integral peak area over the photon energy, Figure 2 shows the resonance spectrum of the respective xenon charge state Xe$^{q+}$.

For the lower xenon charge states $q$=2–6 in Figure 2(a), a broad distribution can be observed, which follows in general the shape of the 4d 'giant' resonance of neutral xenon [16]. Reaching higher charge states ($q$=8,10,15) in Figure 2(b), the width of the charge distribution shrinks and shifts to higher excitation energy for higher charge states.

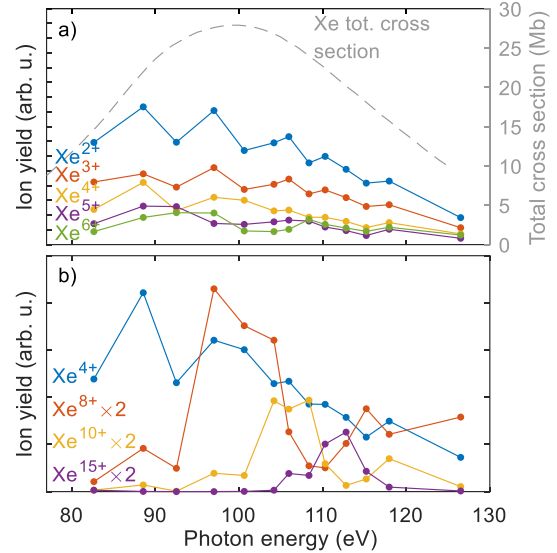

**Figure 2:** Yields of the xenon charge states (a) $q$=2–6 and (b) $q$=8, 10 and 15 as a function of the FEL photon energy for a fluence of 53 J/cm². In (a) cross section of the 'giant' resonance of neutral xenon (grey dashed line) is plotted for comparison [16]. The connecting lines serve for a better overview.

For selected fluences (13 J/cm², 53 J/cm² and 93 J/cm²) and excitation energies (92.5 eV, 104.2 eV, 112.8 eV and 118.0 eV) the matrix in Figure 3 shows the evolution of the ion spectra with fluences compared to changes with the photon energy. The spectra are normalized to Xe$^{2+}$ as Xe$^{2+}$ is predominantly produced by a one-photon process at this photon energy. They show the expected shift towards higher charge states with increasing fluence, but the changes of the ion yields with photon energy are much more pronounced. The yields of the highest charge states are strongly enhanced at 112.8 eV compared to the other excitation energies. Most striking is the difference to the spectra taken at a photon energy of 92.5 eV, which is close to the excitation energy used by Sorokin *et al.*

[3]. Even at the highest fluence shown in Figure 3 of 93 J/cm² (right row) the yield of charge states above Xe$^{7+}$ is significantly reduced compared to the spectrum resulting from 112.8 eV photon energy at a fluence of 13 J/cm² (left row). Thus, the change from 92.5 eV to 112.8 eV photon energy has significantly more impact on the yields of highly charged xenon ions than the increase of fluence by nearly one order of magnitude. In comparison to the data of Sorokin *et al.* [3] the reduced yields of high charge states in our measurement at 92.5 eV despite similar fluences are attributed to the different irradiance as discussed later in the text.

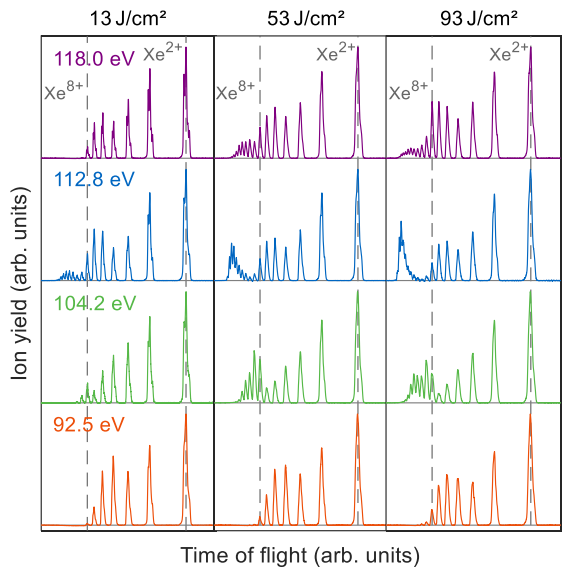

**Figure 3:** Comparison of the xenon photoionization for four selected wavelengths and three photon fluences. All spectra are normalized to Xe$^{2+}$.

The photon energy dependent effect can be quantified by estimating the total absorbed energy of a single xenon atom depicted in Figure 4 for the different fluences (a) 13 J/cm², (b) 53 J/cm² and (c) 93 J/cm². The values were calculated from the identified highest charge state (gray numbers) in every respective spectrum and the ionization potentials from the literature [29].

At 13 J/cm² the highest Xe charge state for excitation energies above 118.0 eV and below 93 eV is Xe$^{10+}$ or lower. In contrast, high charge states of up to Xe$^{17+}$ are already reached at a fluence of 13 J/cm² at 112.8 eV photon energy, which resulted from the more than 3-fold more absorbed total energy compared to 118.0 eV. Even at the fluence of 93 J/cm² the highest charge does not exceed Xe$^{14+}$ in spectra taken with excitation energies at 100.6 eV and below.

The results clearly show the significantly increased efficiency by exciting xenon with photon energies around 110 eV and a respectively strong excitation energy dependency in the ion yield generation of high charge states.

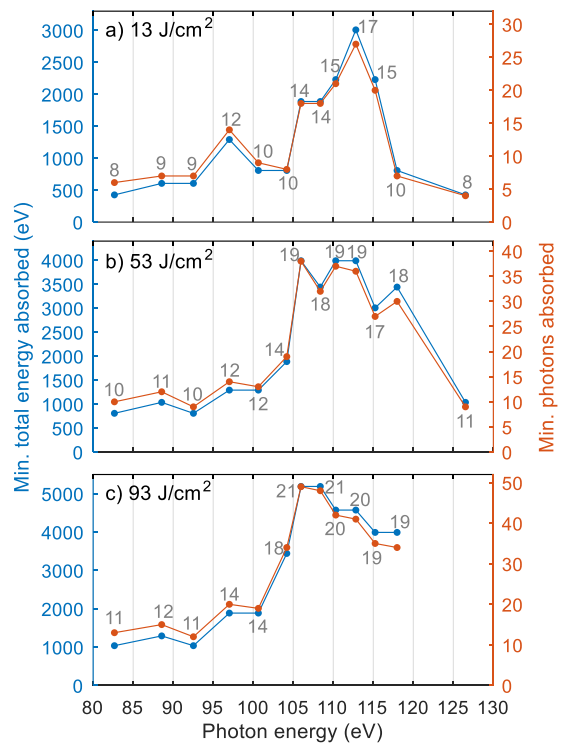

**Figure 4:** Minimal total absorbed energy (blue), minimal absorbed photons (red) and identified highest charge state (gray numbers) for a fluence of (a) 13 J/cm², (b) 53 J/cm² and (c) 93 J/cm². The energy values are the sums of all consecutive ground state ionization thresholds [29].

## IV. Discussion

For photon energies above 106 eV, xenon and its ions in their ground state can be ionized by the absorption of a single photon per ionization step up to $Xe^{8+}$. To ionize the first $4d$ electron in $Xe^{8+}$ from its ground state [Kr] $4d^{10}$ with a completely empty O-shell already 180 eV are required. As the sequential ionization of neutral Xe till $Xe^{8+}$ is already well understood [3, 18, 20, 30-32], we will focus our discussion on ionization processes above $Xe^{8+}$. Starting from $Xe^{8+}$ more than one photon per ionization step is required within the excitation energy range in our experiment and under the assumption of a ground state scenario, i.e., neglecting the population of electronically excited states in previous steps. Earlier theoretical studies for the formation of charge states above $q=8$ showed the continuation of the sequential ionization by multi-photon processes [19, 20] but the mechanism itself has not been revealed up to now.

Simultaneous absorption of two XUV photons, not to mention 7, is several orders of magnitudes less probable than single-photon absorption [10, 33]. In accordance with this, our new findings on the very strong excitation-energy dependency for the yield of highly charge xenon ions support resonant sequential single-photon absorptions involving different pathways and intermediate states.

For the Xe inner-shell excitation using X-rays with a photon energy of 1.5 keV, the influence of the resonant excitation of M-shell electrons and a following Auger-Meitner decay has been discussed by Rudek *et al.* [9]. Ions of higher charge state than $Xe^{17+}$ cannot be ionized directly by a photon of 1.5 keV but it excites resonantly $3l$ electrons into bound $nl'$ states. These states can be further excited or ionized by the next absorbed photon to overcome the ionization threshold until $Xe^{36+}$. In contrast, at a higher photon energy of 2 keV the maximum charge state is much lower due to the unavailability of resonant excitations.

Similar conclusions were drawn by Motomura *et al.* for the excitation of L-shell electrons in xenon with short X-ray pulses of 5.0 keV photon energy [34].

The $4d$ excitation in xenon ions with XUV photons is more complex because more photons per ionization step are required to reach high charge states, for example, five to seven photons are required to ionize $Xe^{18+}$ from its ground state within our selected energy range. We will elucidate that the resonant climbing a sequence of single-photon resonances via excited valence states, which are stable over the FEL pulse, causes the strong photon energy dependency of the ion yields in the XUV and explains the absorption of the high number of XUV photons by xenon.

To support this interpretation, the resonance structure of the intermediate electronic configurations of the different xenon ion species was modeled by utilizing the Hartree-Fock method including relativistic corrections and intermediate coupling for the parent configurations as provided by the program package of Cowan [35] using single-particle wave functions. All resulting absorption spectra are calculated by averaging the spectra from all $^{2S+1}L_J$ states of the ground state configuration and are convoluted with a Gaussian of an FWHM of 1 eV to account for the FEL bandwidth of about 1% of the FEL photon energy. As an example, the calculated absorption spectra for the $Xe^{10+}$ $4d^8$ ground state configuration are depicted in Figure 5(a). Due to the open $4d$ shell, $4p$-$4d$, $4d$-$4f$ and $4d$-$5p$ excitations are possible, all of which have a significant cross section in the range between 90 eV and 125 eV. The absolute energy of the calculated resonance structures has a typical accuracy of 1 eV to 2 eV.

The excited states $4p^5 4d^9$, $4p^6 4d^7 5p$ and $4p^6 4d^7 4f$ can decay only by fluorescence and have lifetimes on the order of nanoseconds and are therefore stable on the 150 fs timescale of the FEL pulse duration. These excited states are characterized by several open shells, which result in a large number of $^{2S+1}L_J$ excitation states in the photon energy range from 90 eV to 125 eV as shown by their corresponding $^{2S+1}L_J$ averaged absorption spectra depicted in Figure 5(b). Therefore, they can be further resonantly excited by the absorption of a second photon.

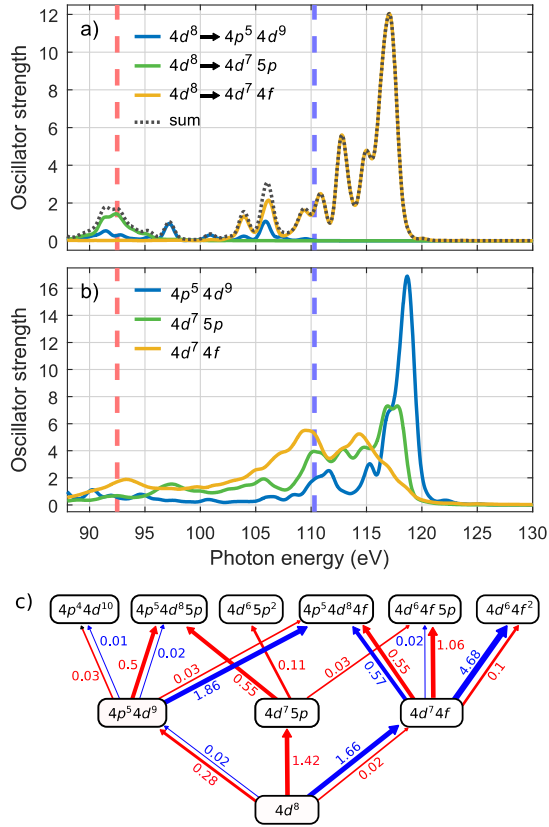

Figure 5: Calculated absorption spectra of the (a) Xe$^{10+}$ $4d^8$ ground state and (b) of the three intermediate states $4p^5\,4d^9$, $4d^7\,5p$ and $4d^7\,4f$ using the Hartree-Fock method provided by the Cowan Code [35] including a convolution with a Gaussian of 1 eV FWHM. Two used FEL photon energies (red: 92.5 eV, blue: 110.3 eV) are marked. (c) Excitation scheme of Xe$^{10+}$ for the first two steps shows the possible transitions and respective oscillator strengths for the two selected excitation energies.

In Figure 5(c), the possible multi-resonant excitation paths from the Xe$^{10+}$ $4d^8$ ground state are sketched for two different excitation energies of 92.5 eV and 110.3 eV, which are marked in parts (a) and (b) with dashed lines. The thickness of the path arrows in Figure 5(c) corresponds qualitatively to the oscillator strengths, which are noted beside the arrows and were taken from the data plotted in Figure 5(a) and (b) at the respective photon energy. For the blue path of 110.3 eV photon energy, the strongest excitations are $4d$-$4f$ transitions, which can result in a strong sequential absorption of two photons and a final $4d^6\,4f^2$ configuration. A $4d^8$ to $4d^7\,5p$ transition is not feasible at 110.3 eV with a calculated oscillator strength in the order of $10^{-28}$. For the red path with 92.5 eV, the $4d$-$5p$ transitions are dominant. Also, $4p$-$4d$ transitions have relatively high oscillator strength. But in contrast to 110.3 eV excitation energy $4d$-$4f$ transitions are unlikely at 92.5 eV.

Depending on the excitation energy, the upper states in Figure 5(c) are already above the ionization threshold of Xe$^{10+}$ of 229 eV and an Auger-Meitner process to the next higher ion charge state becomes possible. If the ionization threshold is not yet surpassed, further resonant excitations can take place and excite the ion further up to the threshold. The strong excitation energy dependency of the xenon ion yields is explained here by these exemplary absorption spectra. While the total oscillator strength is similar for both marked photon energies in the first excitation step in Figure 5(a), further photon absorption and excitation, as can be seen in Figure 5(b), is much more efficient for 110.3 eV. Thus, a series of $4d$-$4f$ transitions, which have high oscillator strengths above 110 eV, result in an increased probability of reaching a state above the ionization threshold. The strong resonances at 118 eV are absent prior to Xe$^{10+}$ (see Figure 6), which reduces the occupancy of high Xe ion charge states at this photon energy.

As shown in the following, these transitions within the N-Shell dominate the ionization process at photon energies between 100 eV and 120 eV for nearly all ion species from Xe$^{8+}$ to Xe$^{20+}$.

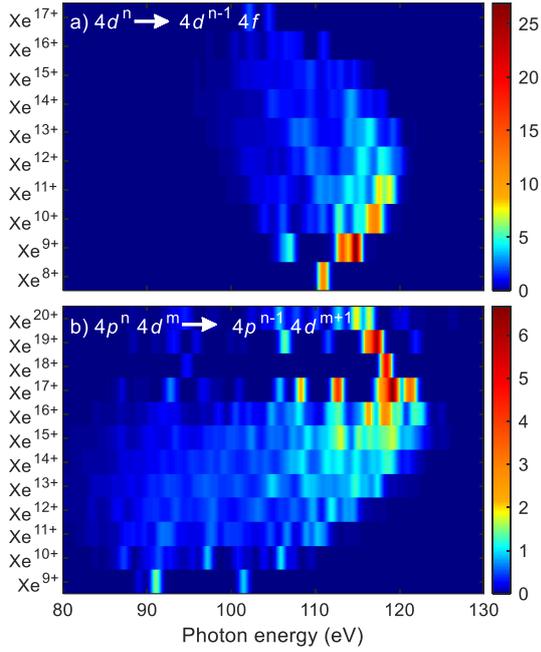

**Figure 6:** False color image of the $^{2S+1}L_J$ averaged calculated absorption spectra for the lowest configuration (a) of $Xe^{8+}$ to $Xe^{17+}$ for $4d$-$4f$ transition and (b) of $Xe^{9+}$ to $Xe^{20+}$ for $4p$-$4d$ transition. The compact depiction simulates classical absorption spectra taken with photographic plates stacked above each other.

We have calculated all N-shell absorption spectra starting at $Xe^{8+}$ for the various ion ladder steps up to $Xe^{20+}$ including $4l$ - $4l'$ and $4l$-$5l'$ transitions. Within the n=4 manifold, all these spectra show strong transitions in the 90 eV to 125 eV excitation energy range. In Figure 6 the calculated $4d$-$4f$ (a) and $4p$-$4d$ (b) $^{2S+1}L_J$ averaged absorption spectra for the lowest configuration of the $Xe^{q+}$ charge states with q=8 to q=20 are depicted in false-color images.

The excitation energies for the $4d$-$4f$ transitions in the differently charged xenon ions are centered around 110 eV with almost no resonance below 100 eV, whereas the weaker $4p$-$4d$ resonances are distributed over a larger energy regime also below 100 eV.

It is obvious that for these higher charge states there always exists a resonance that can be addressed in the 90 eV to 125 eV regime, and that multi-resonant excitations within the n=4 manifold ($4d$-$4f$, $4p$-$4d$) are strongly enhanced. Thus, especially via excitations from $4d$, a highly efficient resonance ladder can be climbed and results in the strong increase of high charge states and energy absorption around 110 eV photon energy. This is well-illustrated by the increasing yields of ions above $Xe^{8+}$ for 106.0 eV to 112.8 eV excitation energy in Figure 1 with $Xe^{16+}$ having the highest yield of all ions above $Xe^{8+}$ at 112.8 eV.

Respectively, we observe a strong decrease in ion yields towards $Xe^{18+}$, when the $4d$ shell is fully depleted. From $Xe^{18+}$ on, the $4p$ resonances can be excited to reach the highest observed charge state of $Xe^{21+}$ within the photon energy range between 90 eV and 125 eV.

By charging xenon from the neutral atom up to $Xe^{18+}$ the binding energies of the n=4 electrons ($4s$, $4p$, $4d$, $4f$) are strongly increasing, but as shown in Figure 6, the transition energies between the different shells with the N-shell stay rather constant. This can be explained by the mean radius of the $4l$ electrons, which would be identical for the hydrogen atom and is changing only slightly here for the case of xenon.

This small variability in the average excitation energy of the resonances for different charge states is special for the $4l$-$4l'$ excitation of xenon, but should also exist in the neighboring elements in the periodic table.

Respectively, possible excitations to Rydberg orbitals with higher quantum numbers are of minor importance in our study, as their average binding energy varies strongly with the charge state. For example, according to our calculations the $4d$-$5p$ resonances of the $Xe^{10+}$ ground state are centered at about 92 eV, which results in a relevant excitation path at 92.5 eV photon energy as seen in Figure 5. However, for the $Xe^{9+}$ and $Xe^{11+}$ ground state the $4d$-$5p$ resonances are centered at about 84 eV and 101 eV, respectively. Accordingly, their contribution to the excitation of the ground state at 92.5 eV is negligible with oscillator strengths of about $3\times10^{-3}$ and $9\times10^{-3}$ for $Xe^{9+}$ and $Xe^{11+}$, respectively. Thus, at fixed excitation energy, 4*l*-5*l'* resonances are accessible only for very few ions and excitation states. One should note here, that the oscillator strengths in ions are in general strongly enhanced compared to neutral atoms [36], resulting in an even enhanced effect for the higher charge states. The validity of the basic mechanism is not influenced by the exclusion of electron correlations in our simulations. Electron-electron correlations are very important processes in ion creation by intense FEL pulses [11]. They open up further ionization pathways and transitions, but are compatible with the explained resonant absorption processes. However, their inclusion in calculations is beyond current capabilities.

Experimentally, electron spectroscopy and electron-ion coincidence are efficient in identifying specific electronic states, transition and electron correlations (see for example [10, 12, 37]). However, the disentanglement of the high amount of different electronic states and resonances involved in the xenon excitation up to $Xe^{21+}$ with intense XUV pulses is similarly beyond reach for now.

As the last point, we want to discuss the influence of the pulse duration by a comparison to the results of the earlier study by Sorokin *et al.* [3]. Their experimental results were obtained at fluences between 0.025 J/cm$^2$ and 78 J/cm$^2$ which also covers a large fraction of the fluence values in this study from 5 J/cm$^2$ to 92 J/cm$^2$. Sorokin *et al.* observed $Xe^{21+}$ at the highest fluence while the highest charge state in our experiment is $Xe^{11+}$ at 92.5 eV and 92 J/cm$^2$. Our spectrum resembles their spectra from 1.4 J/cm$^2$ to 13 J/cm$^2$.

However, although the two excitation energies of 92.5 eV in this study and 93.2 eV have significant overlap assuming a 1% FEL bandwidth, they will excite different resonances on average and follow different excitation pathways. This can be investigated by a follow-up experiment with finer photon energy sampling and, preferably, smaller spectral FEL bandwidth.

But more significance we attribute to the smaller focal spot size of 2.6 µm (FWHM) [3] and a pulse length of 15 fs [21] in the experiment carried out by Sorokin *et al.*, which results in approximately 1 to 2 orders of magnitude higher irradiance at same pulse fluences.

The higher photon density respectively increases the photon absorption rate, which competes with the core-hole lifetime of a feasible Auger-Meitner decay. Thus, although the ionization threshold of an ion species is already exceeded, further energy is absorbed. The following ionization process might result in highly excited ions of the next charge state which therefore needs fewer photons for further ionization than from its ground state or involves a cascade of Auger-Meitner processes. With this it is possible to skip states with weak or even no resonances at the given excitation energy, which hamper further ionization at smaller irradiances.

A similar process has been revealed for the ionization of the xenon atom to $Xe^{4+}$ by Fushitani *et al.* [37], in which a double 4*d* core-hole state is created by two sequentially absorbed photons although the single 4*d* core-hole state already exceeds the ionization threshold. The relaxation via two Auger-Meitner decays increases the charge state by two without further photon absorption.

A deeper insight into the role of irradiance and pulse durations can be obtained with full control or knowledge of the duration of every single FEL pulse. This is expected to be possible in future experiments, e.g. by the realization of online single-shot pulse duration measurement envisaged within the FLASH2020+ project. [38]

## V. Conclusion

By scanning the photon energy along the xenon 'giant' resonance using the variable gap undulators at the FLASH2 facility, we demonstrated a strong variation of the recorded xenon ion yield distribution and the highest identified charge state with the excitation energy. Supported by calculated single-photon absorption spectra using the Cowan Code, we attribute the photon energy dependency to sequential single-photon excitations across several intermediate states.

The mechanism to reach very high charge states in the vicinity of the $4d$ 'giant' resonance using XUV light is, therefore, more complex than the corresponding process at the $L$ or $M$-edges. Only by multi-resonant $4d$-$4f$ excitations within and across the xenon charge states $Xe^{8+}$ to $Xe^{18+}$ one can climb an effective resonance ladder and deposit up 2.6 keV energy in a xenon atom using photon pulses of around 113 eV photon energy already at rather low fluences of 13 J/cm$^2$. We have furthermore revealed that the various open shells in excited xenon create different ladders at slightly different photon energies.

Alongside the two existing models of atomic $4d$ plasma excitations [21] and sequential ionization in the outer shells within perturbation theory [19, 20] we have thereby presented a third explanatory approach for the path towards $Xe^{21+}$ at 93 eV photon energy via a multi-resonant ladder of predominantly $4p$-$4d$ excitations.

This work is one further good example of the absolute necessity to further improve the control and knowledge of experimental conditions and parameter, for example by variable gap undulators and complete shot-to-shot pulse characterization at FELs, as well as constantly revisiting long known samples to gain new insights into fundamental physical processes.

## Acknowledgement

We thank our colleagues from FLASH for the excellent support during the beamtime and B. Keitel for fruitful discussions. SK acknowledges the funding of the EUCALL project within the European Union's Horizon 2020 research and innovation program under the grant agreement no. 654220. MMa. MMe and KB acknowledge funding by the Deutsche Forschungsgemeinschaft (DFG, German Research Foundation) – SFB-925/A3 – project 170620586.